# Epidemic spread in weighted networks

G. Yan[1,§]   Tao Zhou[1,2]   JieWang[1]   Zhongqian Fu[1]   Binghong Wang[2]

1. Department of Electronic Science & Technology,
2. Nonlinear Science Center,   Department of Modern Physics,

University of Science & Technology of China, Heifei, Anhui 230026, P.R.China

We study the detailed epidemic spreading process in scale-free networks with weight that denote familiarity between two people or computers. The result shows that spreading velocity reaches a peak quickly then decays representing power-law time behavior, and comparing to non-weighted networks, precise hierarchical dynamics is not found although the nodes with larger strength is preferential to be infected.

PACS numbers: 89.75.-k, -87.23.Ge, 05.70Ln

Recent studies on epidemic spreading in complex networks indicate a particular relevance in the case of networks characterized by complex topologies and very heterogeneous structure[1-3] that in many cases present us with new epidemic propagation scenarios[4-11], such as absence of any epidemic threshold[6], hierarchical spread of epidemic outbreaks[11]. The new scenarios are of practical interest in computer virus diffusion and the spreading of diseases in heterogeneous populations. It also raises new questions on how to protect the networks and find optimal strategies for the deployment of immunization resource [12,13]. However, so far, studies of epidemic spread just focus on non-weighted complex networks, and a detailed epidemic spreading process in weighted networks is still missing while real networks, such as population and Internet, are obviously scale-free and with links' weights that denote familiarity between two people or computers, respectively.

In this letter, we intend to provide a first analysis of the time evolution of epidemic spreading in weighted networks with highly heterogeneous connectivity patterns. The weighted networks used in this paper is modeled by Alain Barrat, Marc Barthelemy, and Alessandro Vespignani (BBV) [14], in which degree, strength and weight distribution represent heavy tails and power-law behaviors.

In order to study the dynamical evolution of epidemic spreading we shall focus on thee susceptible- infected (SI) model in which individuals can be in two discrete states, either susceptible or infected [15]. The total population(the size of the network) $N$ is assumed to be constant and if $S(t)$ and $I(t)$ are the number of susceptible and infected individuals at time $t$, respectively, then $N = S(t) + I(t)$. In weighted networks, we define the infection transmission by the spreading rate,

$$\lambda_{ij} = (\frac{w_{ij}}{w_M})^\alpha, \qquad (1)$$

at which susceptible individual $i$ acquire the infection from an infected neighbor $j$, where $\alpha$ is a constant $0 < \alpha < 1$, $w_{ij}$ is weight of the link connecting node $i$ and node $j$, $w_M$ is the largest value of $w_{ij}$ in the network. Obviously, more familiar two people are (i.e. larger weight), likelier they infect each other.

We start by selecting one node randomly and assume it's infected. This node infect its neighbors according to the rule of Equ(1). Then the infected nodes infect their neighbors, …, repeat until their neighbors are all infected. The spreading velocity is defined as,

$$V_i(t) = \frac{di(t)}{dt} \approx \frac{I(t) - I(t-1)}{N} \quad (2)$$

where *I(t)* and *i(t)* are the number and fraction of infected vertices, respectively. We register the number of newly infected nodes at each time step and report the spreading velocity in Fig.1. Obviously, velocity goes up to a peak quickly in stead of exponentially increase on non-weighted networks[11], leaving us shorter response times to develop control measures. Moreover, what's interesting, velocity decays following power-law after the "peak time". In a future publication, we will explore deep the reason of velocity's power-law behavior.

In order to give a more precise characterization of the epidemic diffusion through the weighted networks, we measure the average strength of newly infected nodes at time *t*, defined as,

$$\overline{S}_{\inf}(t) = \frac{\int s[I_s(t) - I_s(t-1)]ds}{I(t) - I(t-1)} \quad (3)$$

where $I_s(t)$ is the number of infected vertices with strength *S*. In Fig.2 we plot this quantity as a function of time *t*, and the curves show that $\overline{S}_{\inf}(t)$ represents power-law behavior, $\overline{S}_{\inf}(t) \propto t^{-\gamma}$, in stead of clear hierarchical feature on non-weighted networks[11], and obviously the nodes with larger strength is preferential to be infected.

In summary, we have studied epidemic spreading process in weighted scale-free networks. The result shows that spreading velocity $V_i(t)$ and average strength of newly infected vertices $\overline{S}_{\inf}(t)$ represent power-law time behavior, which is different from infection propagation in non-weighted networks. This indicates that structure of networks impacts epidemic propagation. However, so far, the fundamental problems haven't been resolved, such as, what is the ultimate factor impacting spreading velocity, diameter, heterogeneity or grads of degree density? These will be our future research matters.

___________________________________

§ Electronic Mail: russell0123@ustc.edu

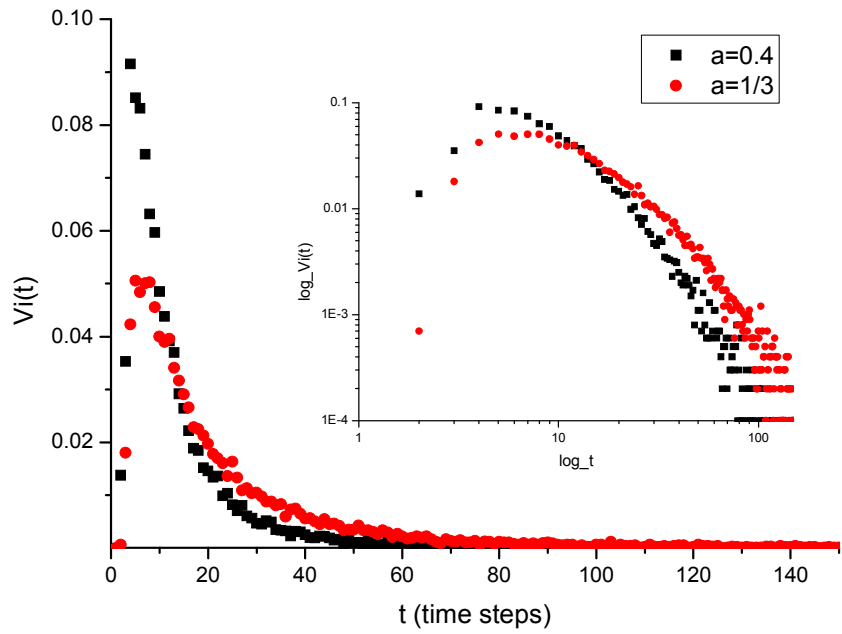

Fig.1 spreading velocity at each time $t$ in weighted scale-free networks proposed by B.B.V. with $N = 10^4, \delta = 3.0, w_0 = 1.0, m = 3$. The inset shows double logarithm curves.

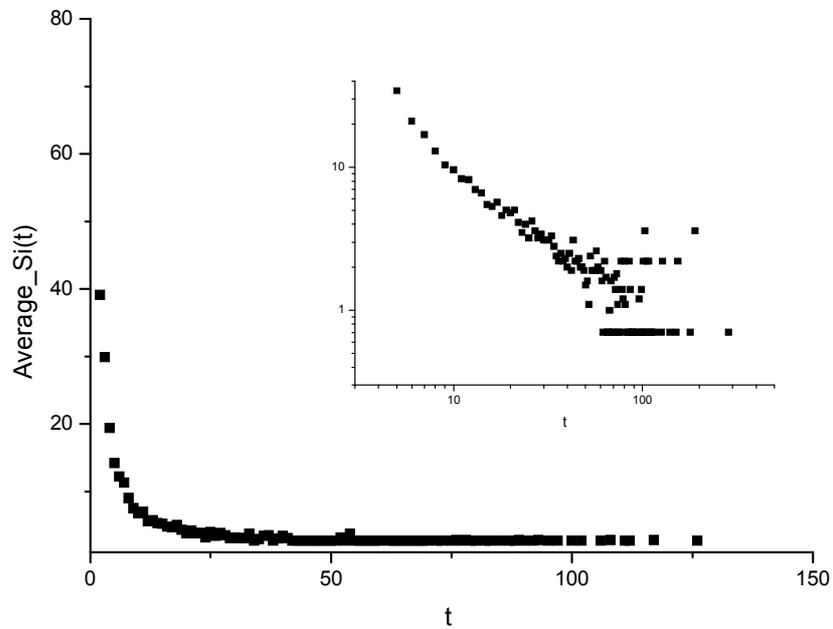

Fig.2 Behavior of average strength of the newly infected nodes at time $t$ for SI model spreading on B.B.V. weighted networks of size $N = 10^4$, the inset shows that $\overline{S}_{inf}(t)$ represents power-law behavior, $\overline{S}_{inf}(t) \propto t^{-\gamma}$.